\documentclass{PoS}
\usepackage{multirow}
\title{Dynamic characterizers of spatiotemporal intermittency}

\ShortTitle{Dynamic characterizers of spatiotemporal intermittency}

\author{\speaker{Neelima Gupte}\\
        Dept. of Physics, Indian Inst. of Technology, Chennai 600 036, India\\
        E-mail: \email{gupte@physics.iitm.ac.in}}

\author{Zahera Jabeen\\
        Dept. of Physics, Indian Inst. of Technology, Chennai 600 036,
India\\
        E-mail: \email{zahera@physics.iitm.ac.in}}

\abstract{
Systems of coupled sine circle maps show regimes of spatiotemporally intermittent behaviour with associated scaling exponents which belong to the DP class, as well as regimes of spatially intermittent behaviour (with associated regular dynamical behaviour) which do not belong to the DP class. Both types of behaviour are seen along the bifurcation boundaries of the synchronized solutions, and contribute distinct signatures to the dynamical characterizers of the system, viz. the distribution of eigenvalues of the one  step stability matrix. Within the spatially intermittent (SI) class, the temporal behaviour of the burst solutions can be quasi-periodic or travelling wave. The usual characterizers of bifurcations, i.e. the eigenvalues of the stability matrix crossing the unit circle, pick up the bifurcation from the synchronized solution to SI with quasi-periodic bursts but are unable to pick up the bifurcation of the synchronized solution to SI with TW bursts. Other characterizers, such as the Shannon entropy of the eigenvalue distribution, and the rate of change of the largest eigenvalue with parameter are required to pick up this bifurcation. This feature has also been seen for other  bifurcations in this system, e.g. that from  the synchronized solution to kink solutions. We therefore conjecture that in the case of high dimensional systems, entropic characterizers provide better signatures of bifurcations from one ordered solution to another.
          }

\FullConference{International conference on Statistical Mechanics of Plasticity and Related Instabilities\\
		 August 29-September 2, 2005\\
		 Indian Institute of Science, Bangalore, India}

\begin{document}

\section{Introduction}

Spatiotemporal intermittency, where regions of laminar or regular behaviour coexist with regions of turbulent or burst behaviour is observed in a wide variety of theoretical and experimental systems\cite{stiex,stith}. In the case of CML-s, the conjecture that spatiotemporal intermittency falls in the same class as directed percolation, has been the subject of extensive investigations \cite{chate, Grassberger, Rolf, bohr, Janaki}. Studies of the diffusively coupled sine circle map show regimes of spatiotemporal intermittency (STI) with associated exponents which exhibit DP behaviour, as well as non-DP regimes of spatial intermittency (SI) where the laminar regions are interspersed with burst regions whose temporal behaviour is periodic or quasi-periodic \cite{zjng}. 

The regimes of spatiotemporal intermittency, as well as those of spatial intermittency lie near the bifurcation boundaries of the synchronized solutions. Each kind of intermittency contributes its distinct signature to the dynamical characterizers of the system, viz. the eigenvalue spectrum of the one step stability matrix. The eigenvalue spectrum is continuous in the case of the STI (DP regime), whereas it shows the existence of gaps in the SI (non-DP) regime \cite{zahera}. 

Both types of intermittency are seen in the neighbourhood of bifurcations from the synchronized solution. In the case of bifurcations from the synchronized solutions to spatiotemporal intermittency, usual stability analysis picks up the transition as the eigenvalues of the stability matrix cross the unit circle. In the case of spatial intermittency, two types of burst states  are seen. The laminar states are synchronized in both cases, but the burst states are quasi-periodic in one case and travelling wave states in the other. The transition from the synchronized state to the spatially intermittent state with quasi-periodic bursts is signalled by the eigenvalues of the stability matrix crossing one, but the transition from the synchronized state to spatial intermittency with the travelling wave bursts is not signalled by the eigenvalues crossing the unit circle. Instead, we see the signature of this bifurcation in other characterizers, such as the Shannon entropy of the eigenvalue distribution, and the rate of change of the largest eigenvalue. This feature has also been seen for other bifurcations in this system, e.g. that from  the synchronized solution to kink solutions. We therefore conjecture that in the case of high dimensional systems, entropic characterizers provide better signatures of bifurcations from one ordered solution to another.

\section{The Model and exponents}
                                                              
\begin{figure}[!t]
\vspace{-.1in}
\begin{center}
\includegraphics[height=2.7in,width=3.8in]{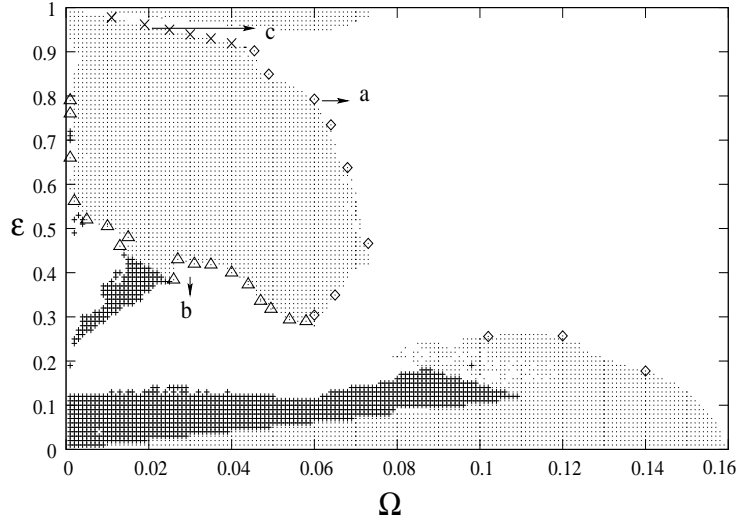}
\vspace{-.1in}
\caption{ shows the phase diagram for the coupled sine circle map lattice using random initial conditions. The spatiotemporally synchronized solutions are represented by dots. The points at which DP exponents have been obtained are marked by diamonds ($\Diamond$). At points marked by triangles ($\triangle$), SI with quasiperiodic bursts is seen. SI with TW bursts is seen at points marked by crosses ($\times$). The cluster solutions are marked as +. The space-time plots obtained at points marked a, b, and c are shown in Figure 2.}
\vspace{-.2in}
\label{pd}
\end{center}
\end{figure}

\begin{figure}
\centering
\vspace{-.35in}
\begin{tabular}{c}
(a)\\

\vspace{-1.0cm}
{
\includegraphics[scale=.65]{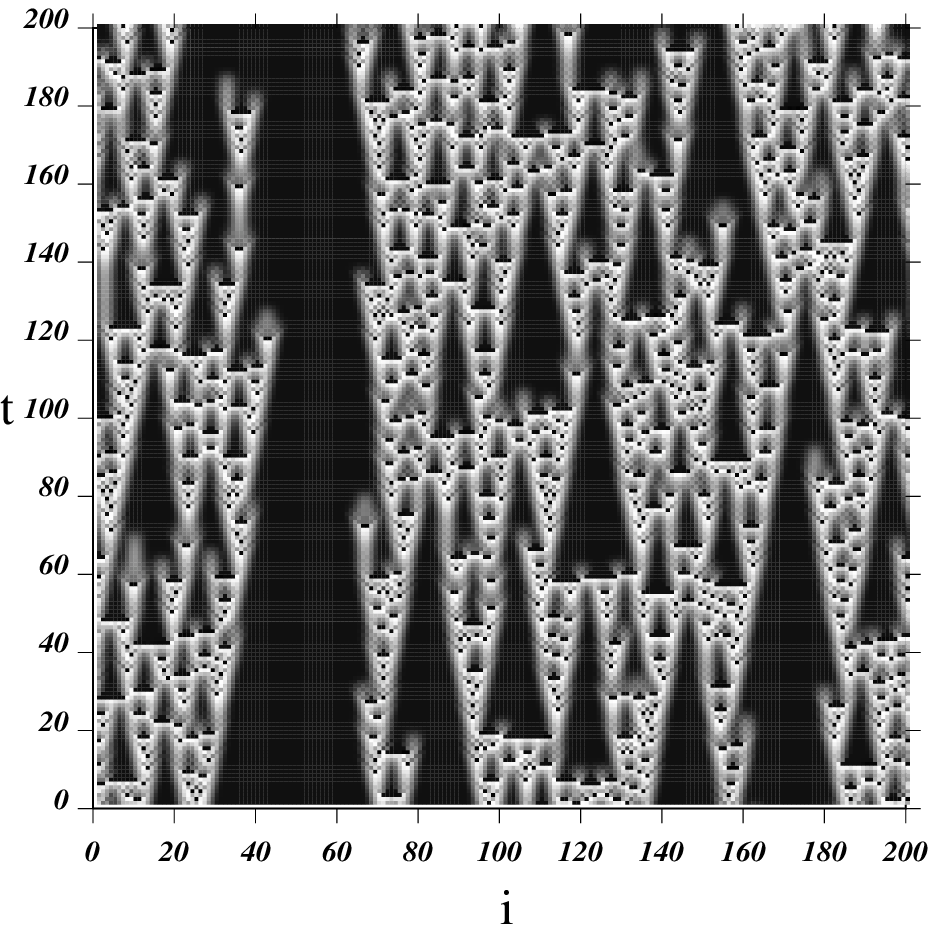}
\vspace{-.70cm}
}\\
\end{tabular}
\begin{tabular}{cc}
(b)&(c)\\
\vspace{-1.cm} 
{\hspace{-.5cm}\includegraphics[scale=.66]{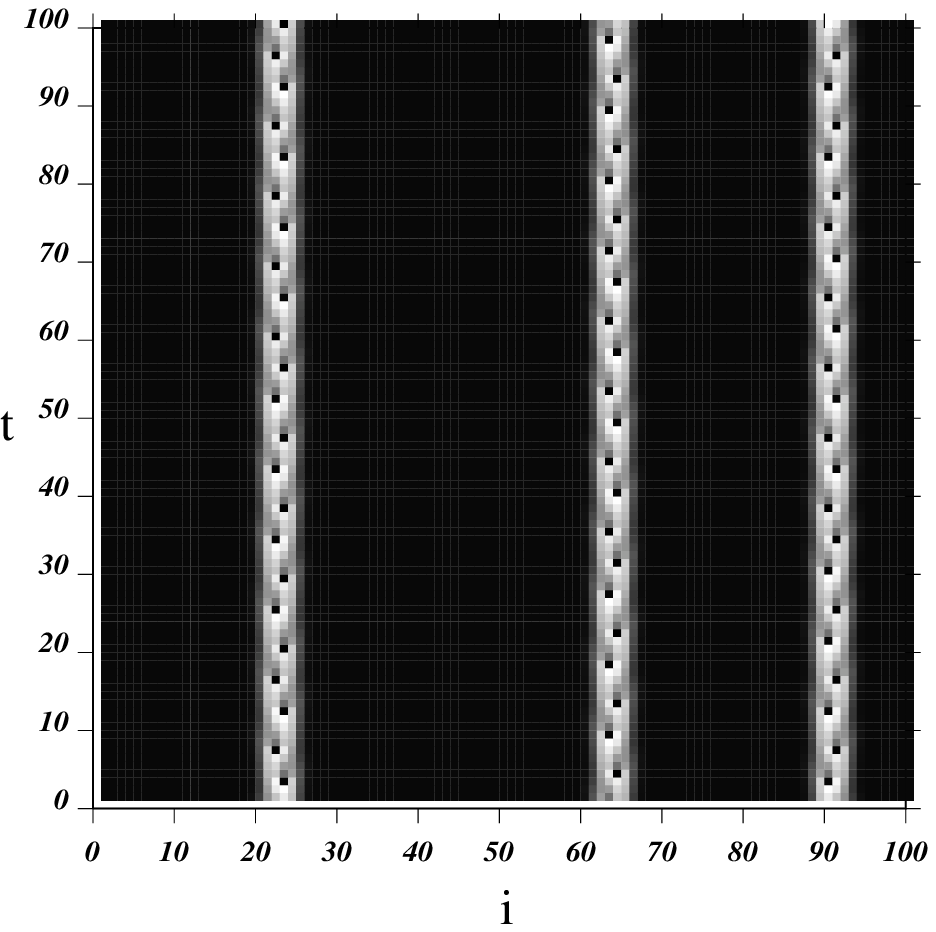}
}&
{\hspace{-.5cm}\includegraphics[scale=.57]{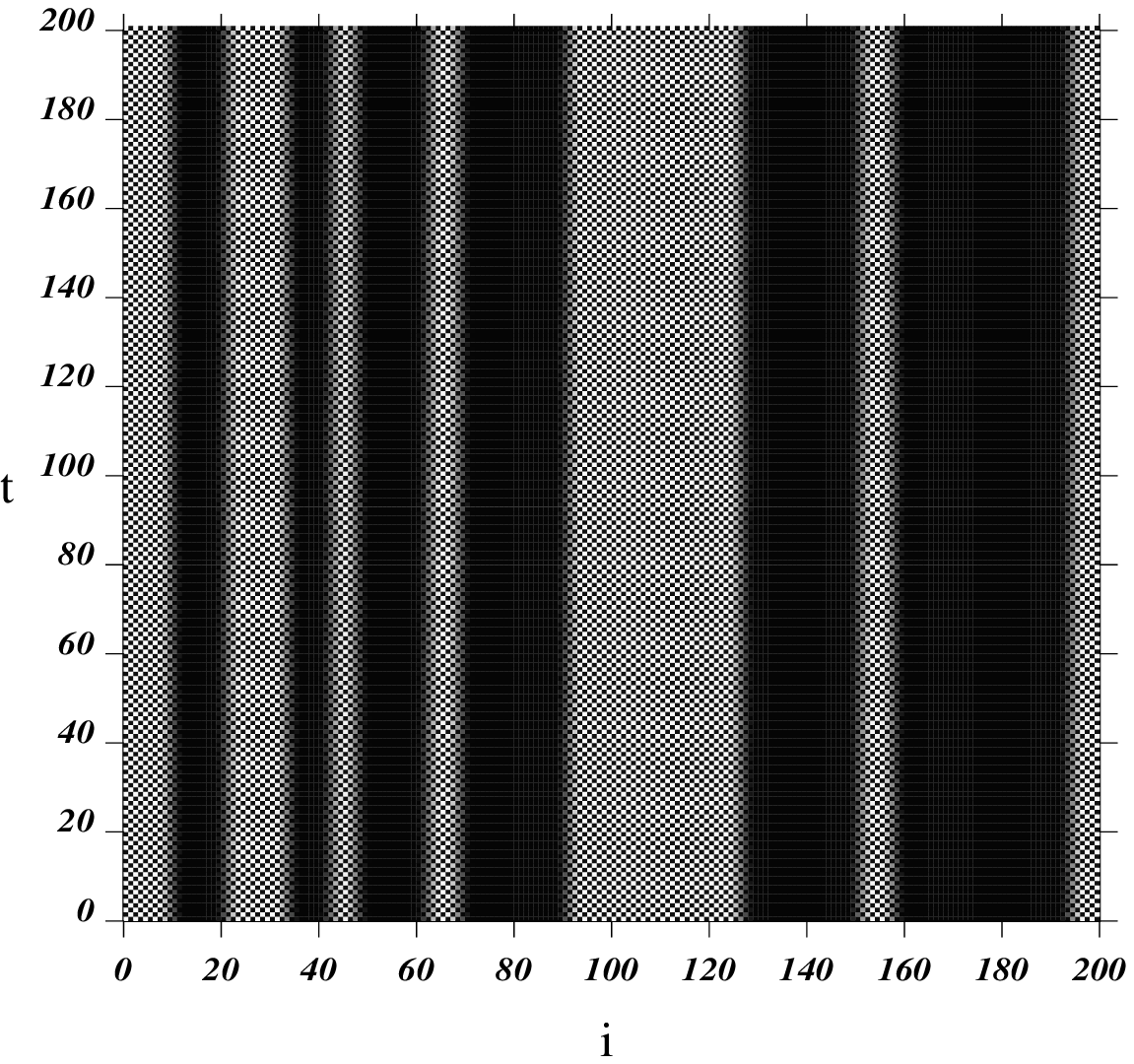}
}
\end{tabular}
\vspace{-.8cm}
\caption{ shows the space time plots of the different types of STI observed in the phase diagram. The lattice index $i$ is along the x-axis and the time index $t$ is along the y-axis. The space time plots show (a) STI with synchronized laminar state interspersed with turbulent bursts seen at $\Omega=0.06, \epsilon=0.7928$. (b) SI with synchronized laminar state with quasi-periodic bursts seen at $\Omega=0.031, \epsilon=0.42$. (c) SI with synchronized laminar state and TW bursts seen at $\Omega=0.019, \epsilon=0.9616$. \label{stplot}}
\end{figure}

\begin{figure}
\centering
%\vspace{-.5in}
\begin{tabular}{c}

(a)\\
{\includegraphics[height=6.cm,width=7.5cm]{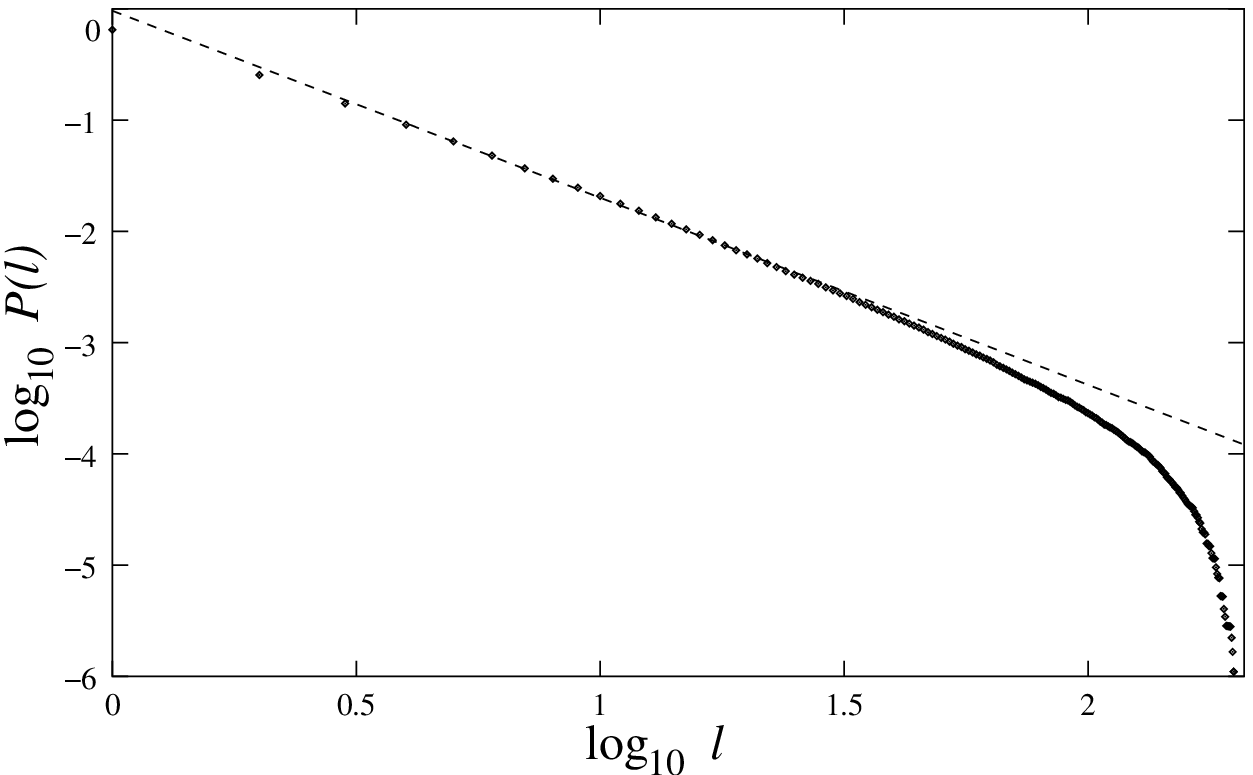}}\\

\end{tabular}

\begin{tabular}{cc}

(b)&(c)\\
{\includegraphics[height=6.cm,width=7.5cm]{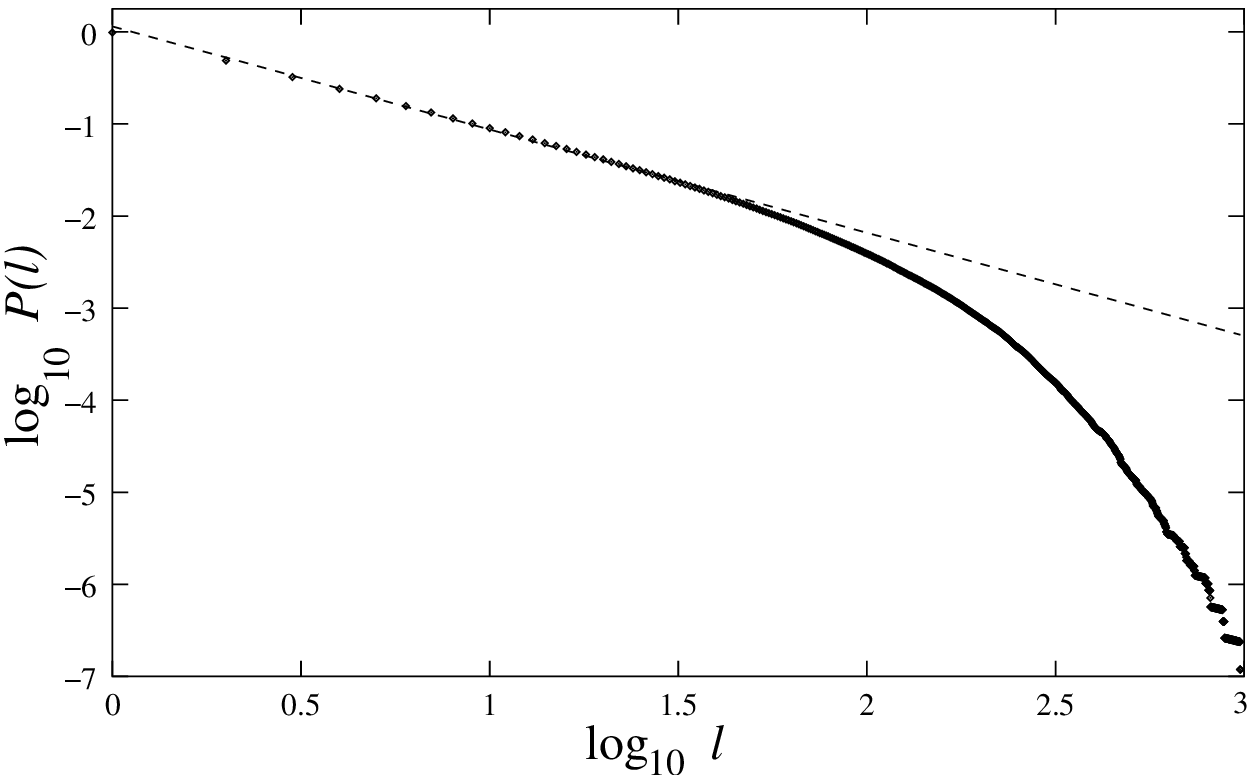}}& {\includegraphics[height=6cm,width=7.5cm]{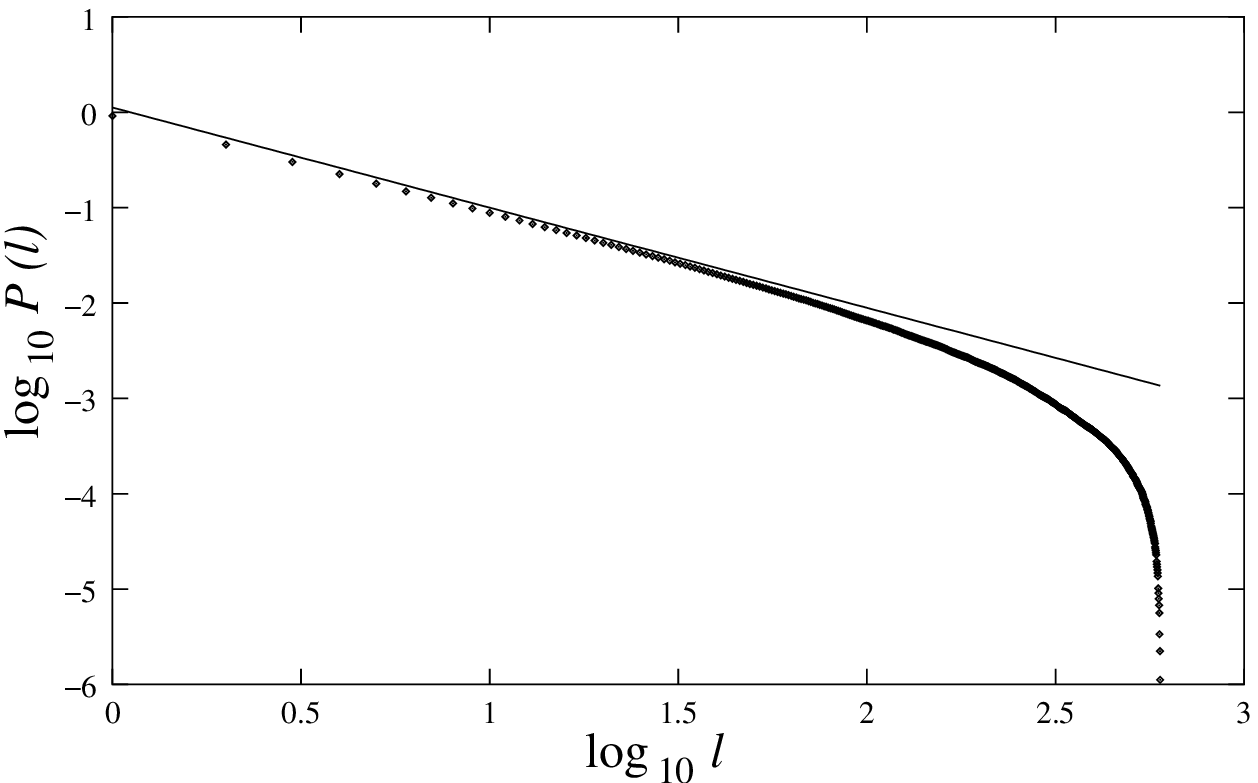}}

\end{tabular}
\caption{shows the $\log-\log$ (base 10) plot of the laminar length distribution for (a)  STI of the DP class obtained at $\Omega=0.06,\epsilon=0.7928$. The exponent $\zeta$ is 1.681. (b) SI with TW bursts obtained at $\Omega=0.019, \epsilon=0.9616$. The exponent obtained is 1.05. (c) SI with quasi-periodic bursts obtained at $\Omega=0.04, \epsilon=0.4$. The exponent $\zeta$ is 1.12. \label{lam}}
\end{figure}

The coupled sine circle map lattice has been known to model the mode-locking behaviour \cite{Gauri1} seen commonly in coupled oscillators, Josephson Junction arrays, etc, and is also found to be amenable to analytical studies \cite{Nandini}. The model is defined by the evolution equations                                       
\begin{equation}
x_i^{t+1}=(1-\epsilon)f(x_i^t)+\frac{\epsilon}{2}[ f(x_{i-1}^t) +
f(x_{i+1}^t) ]\pmod{1}                                                      
\label{evol}
\end{equation}

where $i$ and $t$ are the discrete site and time indices respectively and $\epsilon$ is the strength of the coupling between the site $i$ and its two nearest neighbours. The local on-site map, $f(x)$ is the sine circle map defined as 
\begin{equation}                                                          
f(x)=x+\Omega-\frac{K}{2\pi}\sin(2\pi x)                      
\label{sine}                                                  
\end{equation}                                                          
Here, $K$ is the strength of the nonlinearity and $\Omega$ is the winding number of the  single sine circle map in the absence of the nonlinearity. We study the system with periodic boundary conditions in the parameter regime  $0 < \Omega < \frac{1}{2\pi}$ (where the single sine circle map has temporal period 1 solutions), $0 < \epsilon< 1$ and $K=1.0$. The detailed phase diagram of this model evolved with random initial conditions has been obtained \cite{Janaki, zjng}  and is shown in Fig \ref{pd}. This phase diagram shows regimes of spatiotemporal intermittency (Fig. \ref{stplot}(a)) with accompanying exponents  of the directed percolation type \cite{zahera}, as well as those of spatial intermittency where the burst regions show quasi-periodic or periodic behaviour (Fig. \ref{stplot}(b) and Fig. \ref{stplot}(c)). We discuss behaviour seen at a generic point of each kind (DP and non-DP) in the phase diagram. The generic point chosen for DP behaviour has parameter values $\Omega= 0.06$, $\epsilon= 0.7928$ and $K=1$. We choose two points with non-DP behaviour. The SI with quasi-periodic bursts is seen at $\Omega= 0.04$, $\epsilon= 0.4$ and $K=1$, and SI with periodic bursts is seen at the values $\Omega= 0.019$, $\epsilon= 0.9616$ and $K=1$. To confirm that the point at $\Omega= 0.06$, $\epsilon= 0.7928$ and $K=1$ has DP behaviour, we list the values of the exponents seen in Table \ref{dpt} (See \cite{houlrik, delatorre} for the definition of the DP exponents). The laminar length distribution is plotted in Fig. \ref{lam}(a) and has scaling exponent, $\zeta=1.68$ characteristic of DP behaviour. The laminar length distributions for SI with quasi-periodic bursts  and periodic bursts is plotted in Figs. \ref{lam}(b) and \ref{lam}(c). The length scaling exponent is $\zeta\sim 1.1$ and clearly does not belong to the DP class. 
\begin{table}[!b]
\begin{center}
\begin{tabular}{ll|cccccc|ccccc}
\hline
\multirow{2}{*}{ $\Omega$}&\multirow{2}{*}{ $\epsilon_c(\Omega)$}&\multicolumn{6}{c}{ Bulk exponents}	&\multicolumn{3}{|c}{Spreading Exponents}\\
\cline{3-11}
 & &  $z$ &  $\beta/{\nu z} $ &  $\beta$ &  $\nu$ &  $\eta'$ & { $\zeta$} & $\eta$	&$\delta$	&$z_s$\\
\hline
 	&	&	&	&	&		&	&	&	&\\
0.06	&0.7928	&1.59	&0.17	&0.29	&1.1	&1.51	&1.68&0.315 &0.16	&1.26	\\
\hline
\multicolumn{2}{c|}{\bf DP}& {\bf 1.58}	& {\bf 0.16}	& {\bf 0.28}	& {\bf 1.1}	& {\bf 1.51}	&{\bf 1.67} & {\bf 0.313}	&{\bf 0.16}	&{\bf 1.26}		\\
	
 \hline
\end{tabular}
\caption{ The static and dynamic exponents obtained  at the critical $\epsilon_c$ are shown in the above table. The universal DP exponents are listed in the last row.  \label{dpt}}
\end{center}
\end{table}

\clearpage

\section{Dynamic characterizers}
\subsection{Differences between the DP and non-DP class}
The linear stability matrix of the evolution equation \ref{evol} at one time-step about the  solution of interest is given  by the  $N \times N$ dimensional matrix, $M_t^N$, given below 

\begin{displaymath}
\mathbf{M_t^N} = 
\left( \begin{array}{cccccc}
\epsilon_s f'(x_1^t) & \epsilon_n f'(x_2^t) & 0 & \ldots& 0 &\epsilon_n f'(x_N^t)\\
\epsilon_n f'(x_1^t)&\epsilon_s f'(x_2^t)&\epsilon_n f'(x_3^t)&0&\ldots&0\\
0&\epsilon_n f'(x_2^t)&\epsilon_s f'(x_3^t)&\ldots&0&0\\
\vdots&\vdots&\vdots&\vdots&\vdots&\vdots\\
\epsilon_n f'(x_1^t)&0&\ldots&~~~0~~~&\epsilon_n f'(x_{N-1}^t)&\epsilon_s
f'(x_N^t)\\ 
\end{array}\right)
\end{displaymath}

where, $\epsilon_s=1-\epsilon$, $\epsilon_n=\epsilon/2$, and $f'(x_i^t)=1-K\cos (2\pi x_i^t)$. $x_i^t$ is the state variable at site $i$ at time $t$, and a lattice of $N$ sites is considered. 
                                  
The diagonalisation of $M^N_t$ gives the $N$ eigenvalues of the stability matrix.  The eigenvalues of the stability matrix were calculated for spatiotemporally intermittent solutions which result from bifurcations from the spatiotemporally synchronized solutions (frozen, homogenous solutions). The eigenvalue distribution for all the cases was calculated by averaging over 50 initial conditions.

\begin{figure}[!t]
\begin{center}                                                
\hspace{-.4in}\includegraphics[height=6cm,width=13.1cm]{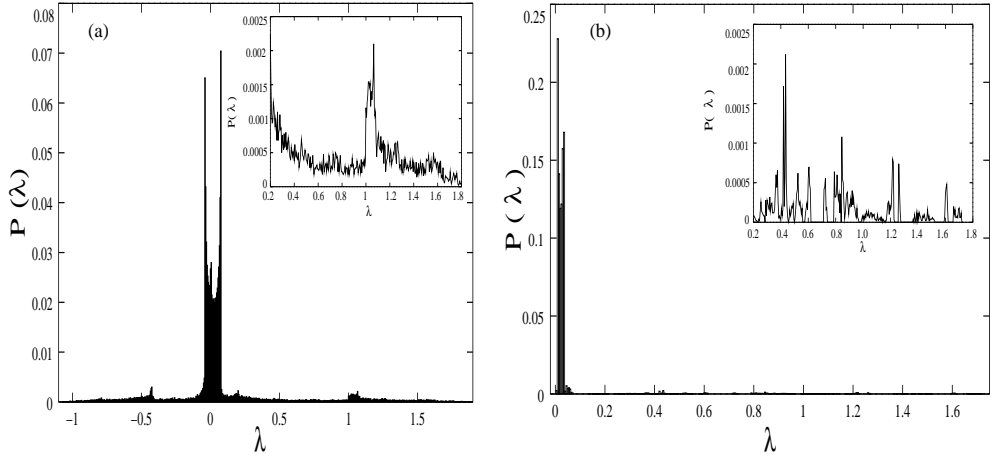}
\caption{ shows the eigenvalue distribution (binsize=0.005) for (a) STI belonging to the DP class at $\Omega=0.06,\epsilon=0.7928$ and (b) Spatial intermittency with quasi-periodic bursts at $\Omega=0.04,\epsilon=0.4$. A section of the eigenvalue distribution is magnified in the inset figures. Gaps are seen in the spatial intermittency eigenvalue distributions whereas the eigenvalue distribution for STI does not show any such gaps.\label{nhgm}}
\vspace{-.2in}
\end{center}                                                  
\end{figure}   
 The eigenvalue distributions for STI belonging to the DP class can be seen in  Fig. \ref{nhgm}(a), and  that for spatial intermittency can be seen in Fig. \ref{nhgm}(b) and Fig. \ref{hg35}. It is clear from the insets that the eigenvalue spectrum of the SI case shows distinct gaps whereas no such gaps are seen in the eigenvalue spectrum of the STI belonging to the DP class and the spectrum is continuous. Thus, a form of level repulsion is seen in the eigenvalue distribution for parameter values which show spatial intermittency. We note that such gaps are seen at all the parameter values studied where spatial intermittency is seen, and that no gaps are seen for any of the parameter values where DP is seen for binsizes of 0.005 and above.
                               
\subsection{Characterization of the transition to spatial intermittency}
\begin{figure}[!t]
\begin{center}
\includegraphics[height=5cm,width=6.5cm]{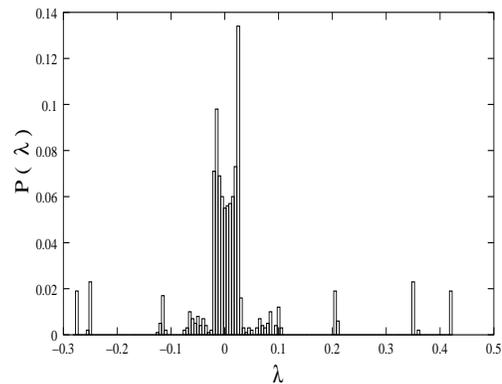}
\end{center}
\caption{ shows the eigenvalue distribution of Spatial intermittency with TW bursts at $\Omega=0.035,\epsilon=0.9294$ obtained for one initial condition. The modulus of the largest eigenvalue is less than one .\label{hg35}}
\end{figure}

In all the cases of spatiotemporal intermittency seen here, the synchronized solutions bifurcate to the intermittent state. It is clear from Figs \ref{nhgm}(a) and \ref{nhgm}(b) that the spatiotemporally intermittent solution and the spatially intermittent solution with quasi-periodic behaviour have bifurcated from the synchronized solution via a tangent-tangent bifurcation, and several eigenvalues have crossed the unit circle at $+1$. However, it is clear that Fig. \ref{hg35} shows different behaviour. The gaps characteristic of SI are seen, however, though the solution has changed from the synchronized solution to spatially intermittent behaviour with travelling wave bursts, the eigenvalues have not crossed the unit circle. Thus, of the two transitions to spatial intermittency, only one is picked up by the usual stability criterion.   

This feature has also been seen for other bifurcations in this system, e.g. that from  the synchronized solution to kink solutions \cite{Gauri2}, and that from synchronized solutions to frozen spatial period two solutions \cite{Gauri1}. It was seen that in these cases too, the bifurcation from the synchronized solution was not picked up by the usual characterizers. It was seen in these cases that the rate of change of the largest eigenvalue as well as the distribution  of eigenvalues were better able to pick up the bifurcation \cite{Gauri2}.

\begin{figure}[!t]
\begin{center}
\vspace{.2in}
\begin{tabular}{cc}
{\includegraphics[height=5cm,width=6.5cm]{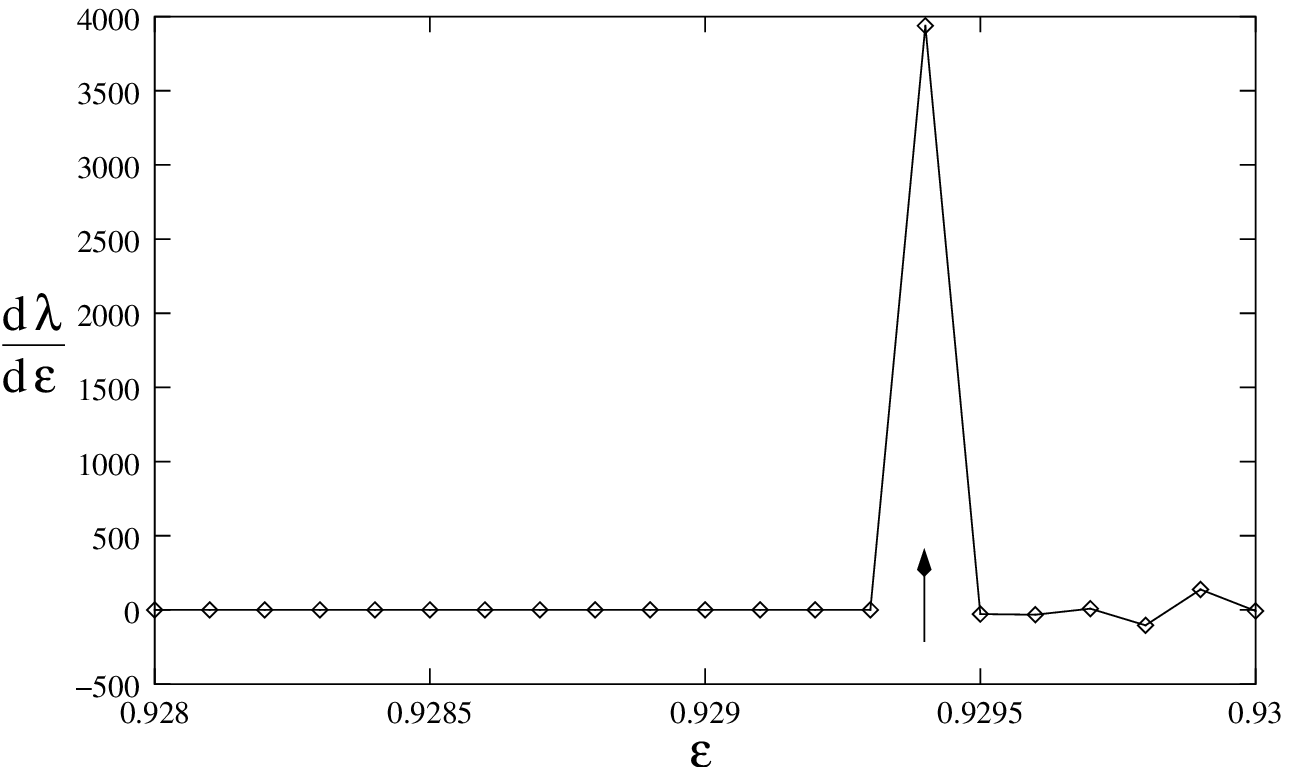}}
&
{\includegraphics[height=5cm,width=6.5cm]{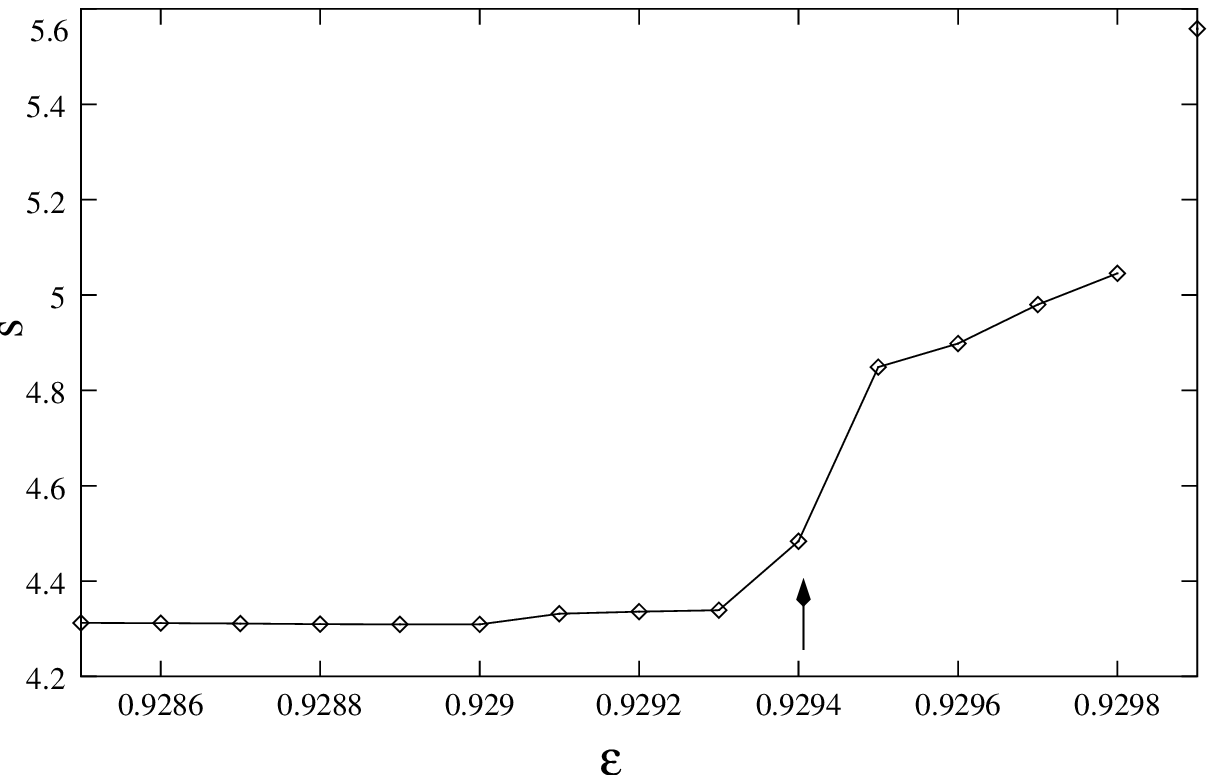}}
\end{tabular}
\end{center}
\caption{ (a) shows $d\lambda/d\epsilon$ plotted against $\epsilon$ at $\Omega=0.035$. A sharp jump is seen at $\epsilon=0.9294$ where the bifurcation has taken place. (b) shows the Shannon entropy, $S$ plotted against $\epsilon$ at $\Omega=0.035$. A jump is seen at $\epsilon=0.9294$.\label{ds}}
\end{figure}

We study the behaviour of the rate of change of the largest eigenvalue $\lambda$ with the parameter $\epsilon$ at fixed $\Omega$ in Fig. \ref{ds}(a). It is clear that the rate of change shows a clear jump at the transition, and the bifurcation is clearly picked up. We also define the  Shannon entropy, $S=-\Sigma_i p_i \log p_i$ as a characterizer of the bifurcation. Here, $p_i$ is the probability that the eigenvalue takes the value corresponding to the box $i$, and the sum is taken over all the occupied boxes. This quantity is plotted as a function of $\epsilon$ in Fig. \ref{ds}(b). It is clear that the bifurcation is picked up by this quantity as well.
\clearpage

\begin{table}[!b]
\begin{tabular}{lllll}
\hline
Type of solution & $\Omega$ & $\epsilon$ & Largest eigenvalues & Bifurcation Type\\
\hline
 & & & & \\
Synchronized, frozen &{0.04} &0.4045& 0.032104, 0.032104 &\\
SI with QP bursts&0.04 &0.4044&1.701346, 1.270010& Double Tangent\\
\hline
 & & & & \\
Synchronized, frozen& {0.035}& 0.9293 & 0.024475, 0.024475&\\
SI with TW bursts & 0.035&0.9294 & 0.418326, 0.415696&{~~~~~~~~~~-}\\
\hline
\end{tabular}

\caption{ shows the largest eigenvalues obtained for SI with quasi-periodic bursts (QP) and SI with TW bursts. (a) The bifurcation to SI with quasi-periodic bursts (at $\Omega=0.04, \epsilon= 0.4044$) from synchronized solutions (at $\epsilon=0.4045$) has taken place through a tangent-tangent bifurcation. (b) Synchronized solutions at $\Omega=0.035, \epsilon=0.9293$ bifurcate to SI with TW bursts at $\epsilon=0.9294$. The two largest eigenvalues have modulus less than one even though the solution has changed. \label{lmax}}
\end{table}

\clearpage
\section{Discussion}

We find that spatiotemporal intermittency in the case of the sine circle map lattice shows both DP and non-DP behaviour depending on the parameter regime. Each type of behaviour contributes its own characteristic exponent to the distribution of laminar lengths. Signatures of DP and non-DP behaviour are found in the dynamic characterizers, viz. the eigenvalue distribution of the one step stability matrix. Within the non-DP class, two types of spatial intermittency are seen, one where the burst behaviour is quasi-periodic and the other where the burst behaviour is periodic. Transitions from the synchronized state to the first type of SI are signalled by the usual stability criterion of the eigenvalues of the stability matrix crossing one, but the other bifurcation, i.e that to the second type of SI is not picked up by the usual criterion. Entropic characterizers are useful for picking up this kind of transition, and the Shannon entropy of the distribution of eigenvalues is sufficient to pick up this bifurcation. Such characterizers may be useful in other contexts as well.

\section{Acknowledgement}
N.G. thanks BRNS, India for partial support. Z.J. thanks CSIR, India for financial support.

\clearpage

\end{document}